\documentclass[9pt,twocolumn,twoside]{pnas-new}
\usepackage{graphicx,subfigure}
\usepackage[flushleft]{threeparttable}
\usepackage{amssymb, amsmath, mathtools}

\templatetype{pnasresearcharticle} 

\title{The artificial ecosystem: number soup (part II)}

\author[]{Yu Liu}

\affil[]{Department of Mathematics, Uppsala University, 75105 Uppsala, Sweden}

\leadauthor{Liu} 

\correspondingauthor{\textsuperscript{}E-mail: yu.liu@math.uu.se}


\begin{abstract}
This paper is a follow-up work about the artificial ecosystem model: number soup (Liu and Sumpter, \textit{J. Royal Soc. Interface}, 2017). It elaborates more details about this model and points out future directions.
\end{abstract}

\begin{document}

\verticaladjustment{-2pt}

\maketitle
\thispagestyle{firststyle}
\ifthenelse{\boolean{shortarticle}}{\ifthenelse{\boolean{singlecolumn}}{\abscontentformatted}{\abscontent}}{}

\dropcap{T}his paper is a follow-up work about the number soup model \cite{Liu17IIR}. It elaborates more mathematical details; describes more general properties and gives different interpretations of the model in a more heuristic way; and lists some intriguing questions deserved to be investigated more.

The paper is organised in the following way. Section \ref{sec:biodiv} shows examples of demographic structures in different $n$-metabolite systems, aiming to give a feeling about how complex the interactions among species and how diverse an ecosystem could be. Section \ref{sec:speciesloop} discusses the properties of the ecological concept we put forward, the species loop, in more detail. Section \ref{sec:Gill} connects Gillespie Algorithm and the intuition that on average one unit time is the duration that each organism completes a life cycle. Section \ref{sec:Monod} compares Monod equation, an classic empirical law for microbial growth, with the growth rate of organisms I assumed in the number soup model. Section \ref{sec:addition} describes another way of thinking the rules we used in the model about what a species excretes. Section \ref{sec:typesCF} clarifies different types of cross-feeding, only one of which is tackled by the number soup model. Section \ref{sec:cheater} tries to interpret the model in the view of cooperators and cheaters, promoting sceptical thinking. Section \ref{sec:future} lists intriguing points and questions which are deserved to be investigated, associated with the number soup model, in the order of relevance and importance.

\section{Rich biodiversity}
\label{sec:biodiv}
 
When $n$, the number of all possible metabolites, is large, the number soup model produces rich biodiversity.

\begin{figure*}[tbhp] \centering
  \subfigure { \label{fig:9metabo1}
    \includegraphics[width = 0.42\linewidth]{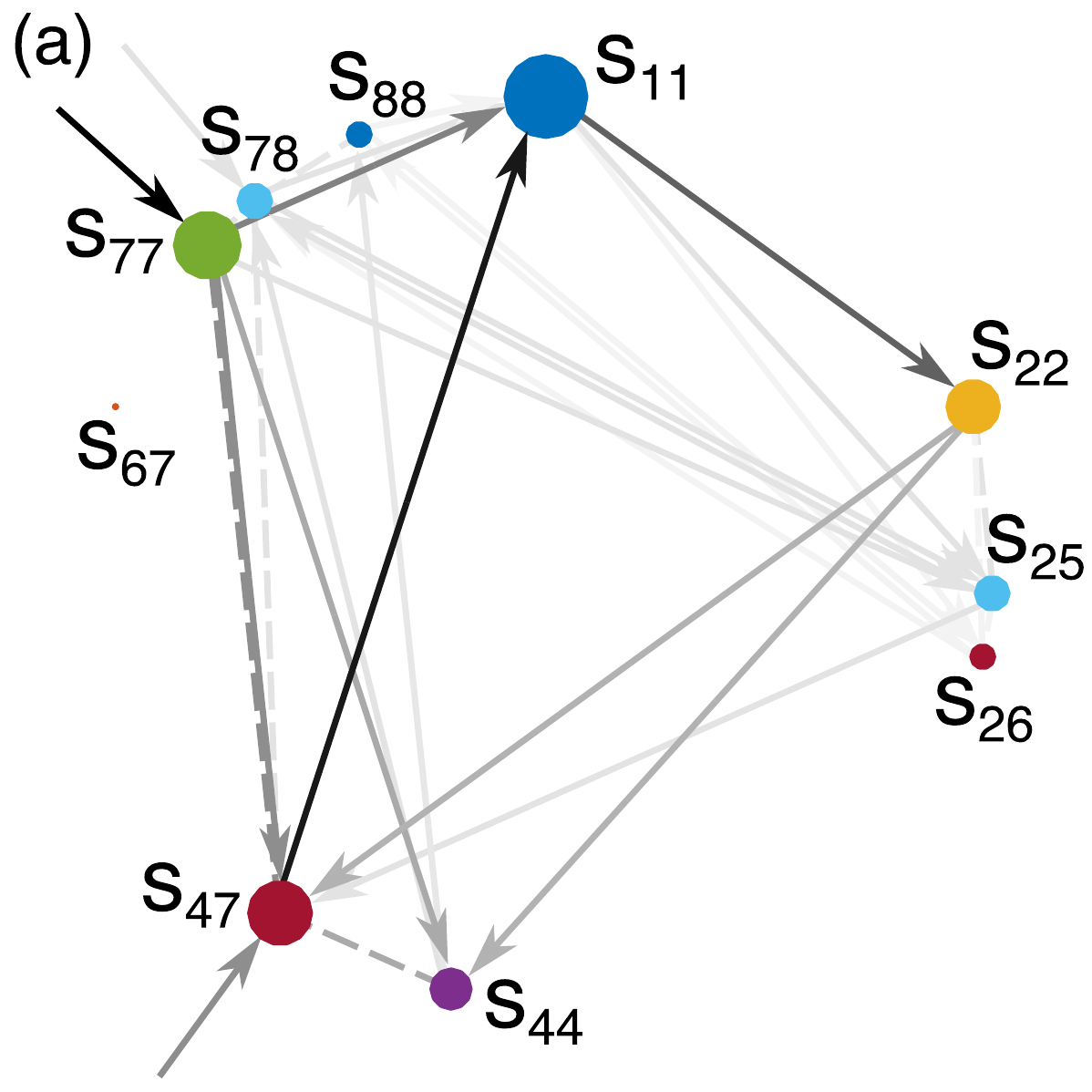}}\hspace*{3em}
  \subfigure { \label{fig:9metabo2}
    \includegraphics[width = 0.42\linewidth]{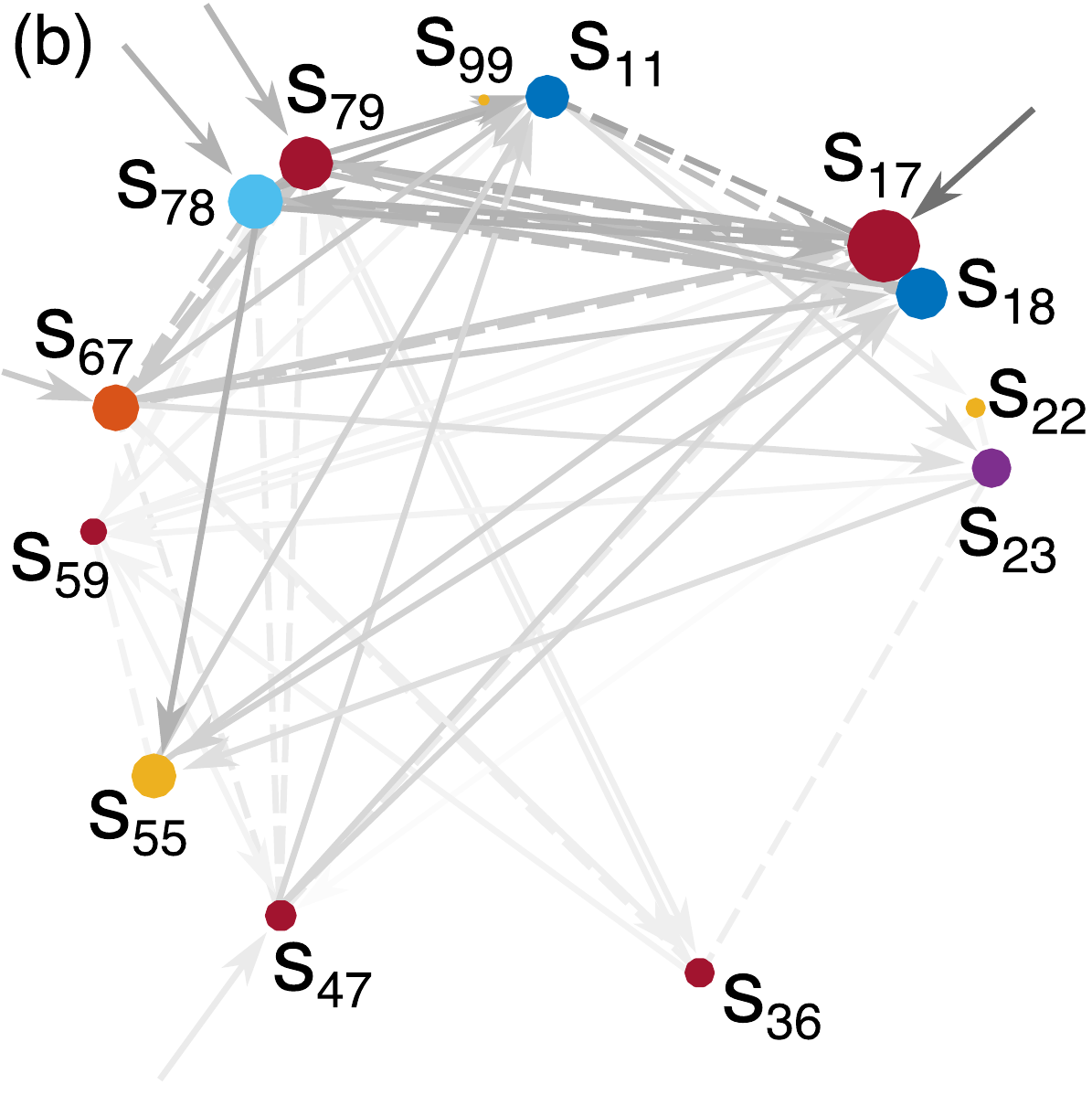}}
  \subfigure { \label{fig:9metabo3}
    \includegraphics[width = 0.42\linewidth]{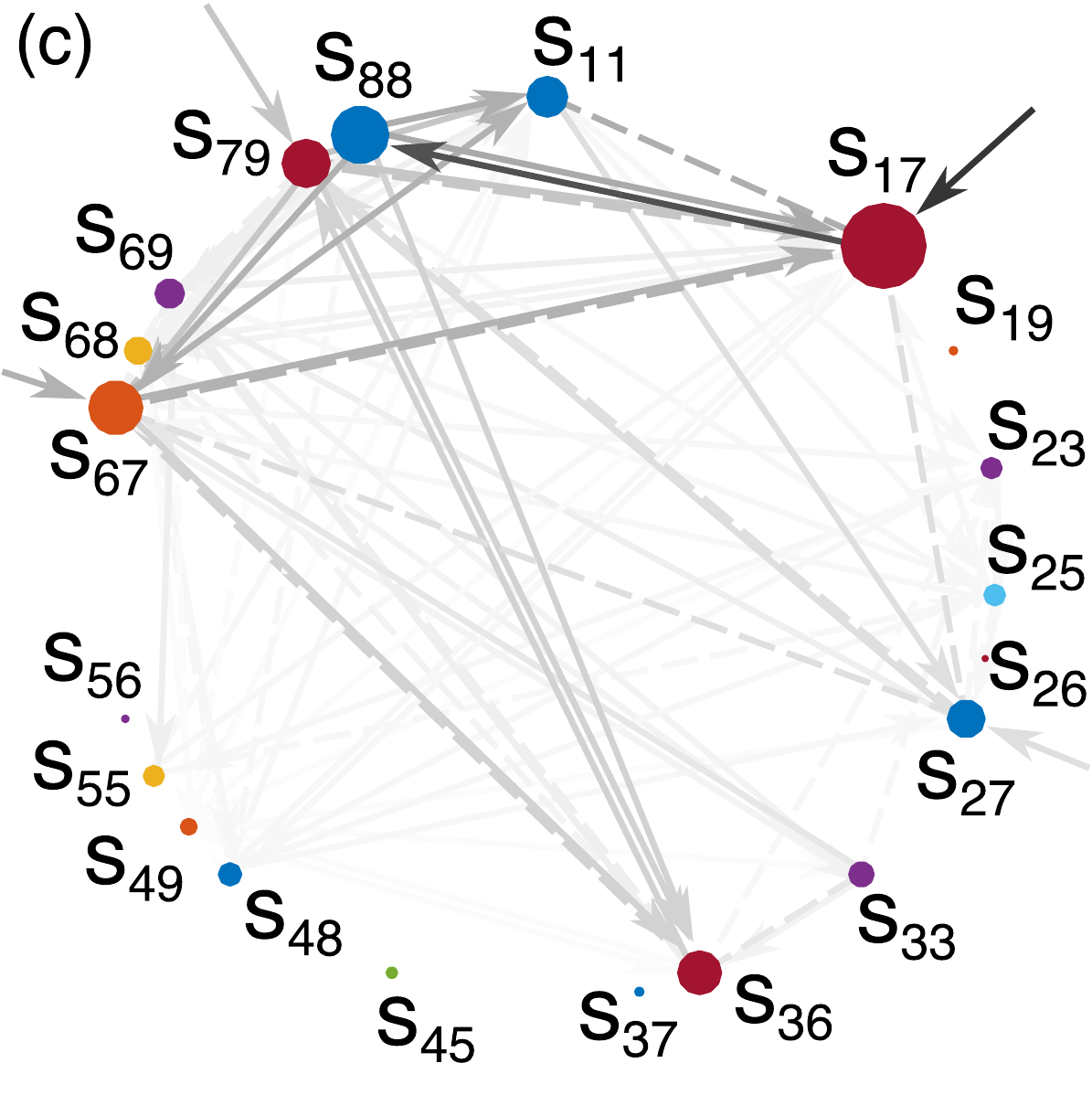}}\hspace*{3em}
  \subfigure { \label{fig:9metabo4}
    \includegraphics[width = 0.42\linewidth]{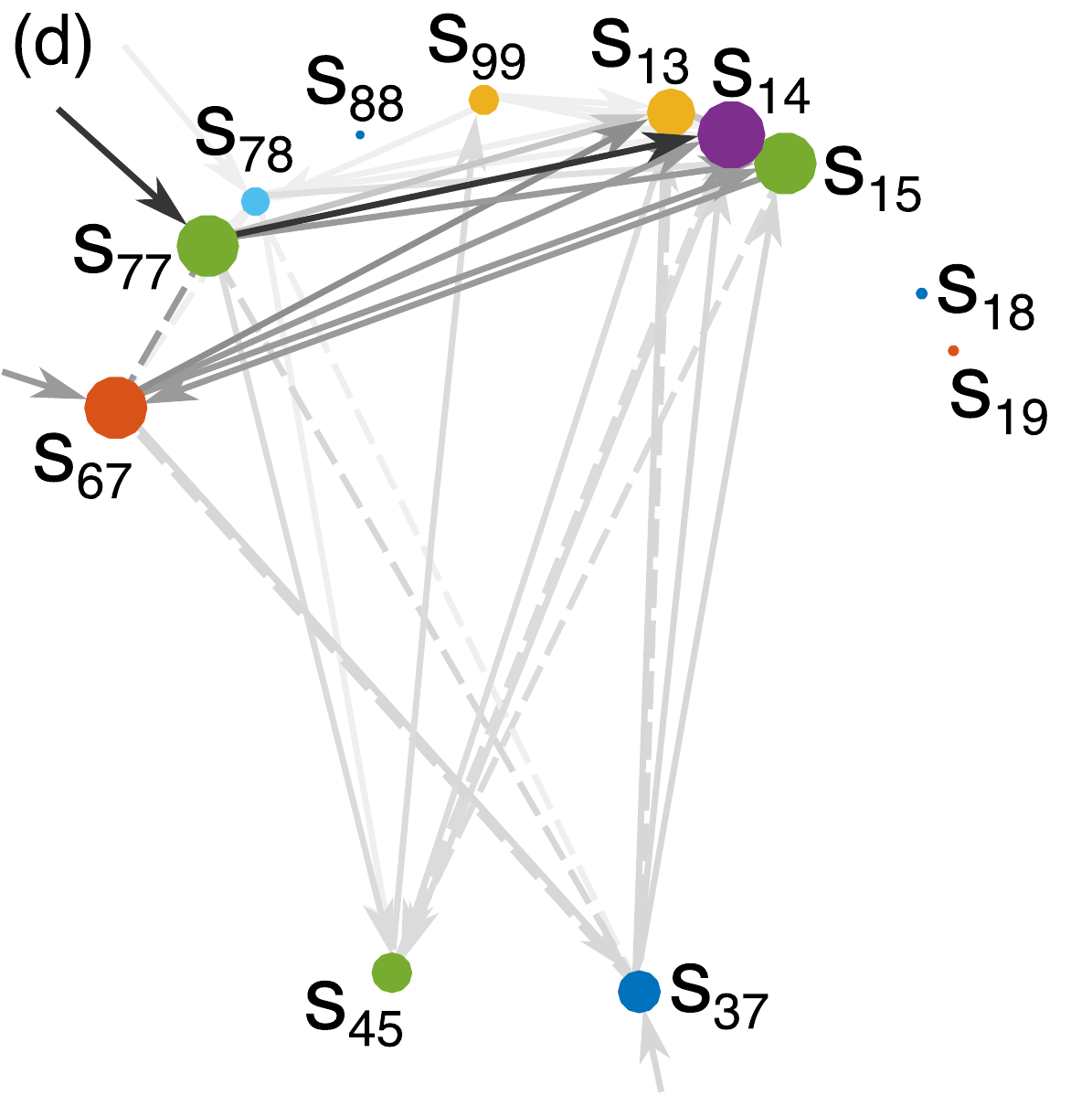}}
\caption{Visualisation of four demographic structures in the 9-metabolite system with inflowing metabolite $\bar{7}$. The positions of all the 45 species are aligned evenly in a circle in the way that $s_{11}, s_{12}, \cdots, s_{18}, s_{22}, s_{23}, \cdots, s_{28}, s_{33}, \cdots, s_{88}$ where $s_{11}$ is in the 12 o'clock position. The species with zero population is not shown in the figures. All the four figures have the same normalisation for the population size and interaction strength.}
\label{fig:9metabo}
\end{figure*}

Figure \ref{fig:9metabo} shows four different demographic structures in the 9-metabolite system with inflowing metabolite $\bar{7}$. The same parameters and initial conditions are used in the four figures, and they are just different runs.
In the demographic structure shown in Fig. \ref{fig:9metabo1}, species $s_{11}, s_{77}, s_{47}$ and $s_{22}$ are among the most abundant. Species $s_{77}$ takes in the inflowing metabolite $\bar{7}$ and produce $\bar{1}$ and $\bar{4}$, which are taken in by species $s_{11}$ and $s_{47}$. And species $s_{11}$ produces $\bar{2}$, being taken in by species $s_{22}$ which produces $\bar{4}$, being taken by $s_{47}$ again. This is the main ``metabolism'' of the ecosystem in this demographic structure. Besides these strong interactions, as we can see, there are more metabolic reactions with weaker interactions involved. For another demographic structure, e.g., Fig. \ref{fig:9metabo2}, totally different species coexist. There are more species coexisting while the interactions are weaker, comparing with that in Fig. \ref{fig:9metabo1}.

\begin{figure*}[tbhp] \centering
  \subfigure { \label{fig:16metabo1}
    \includegraphics[width = 0.48\linewidth]{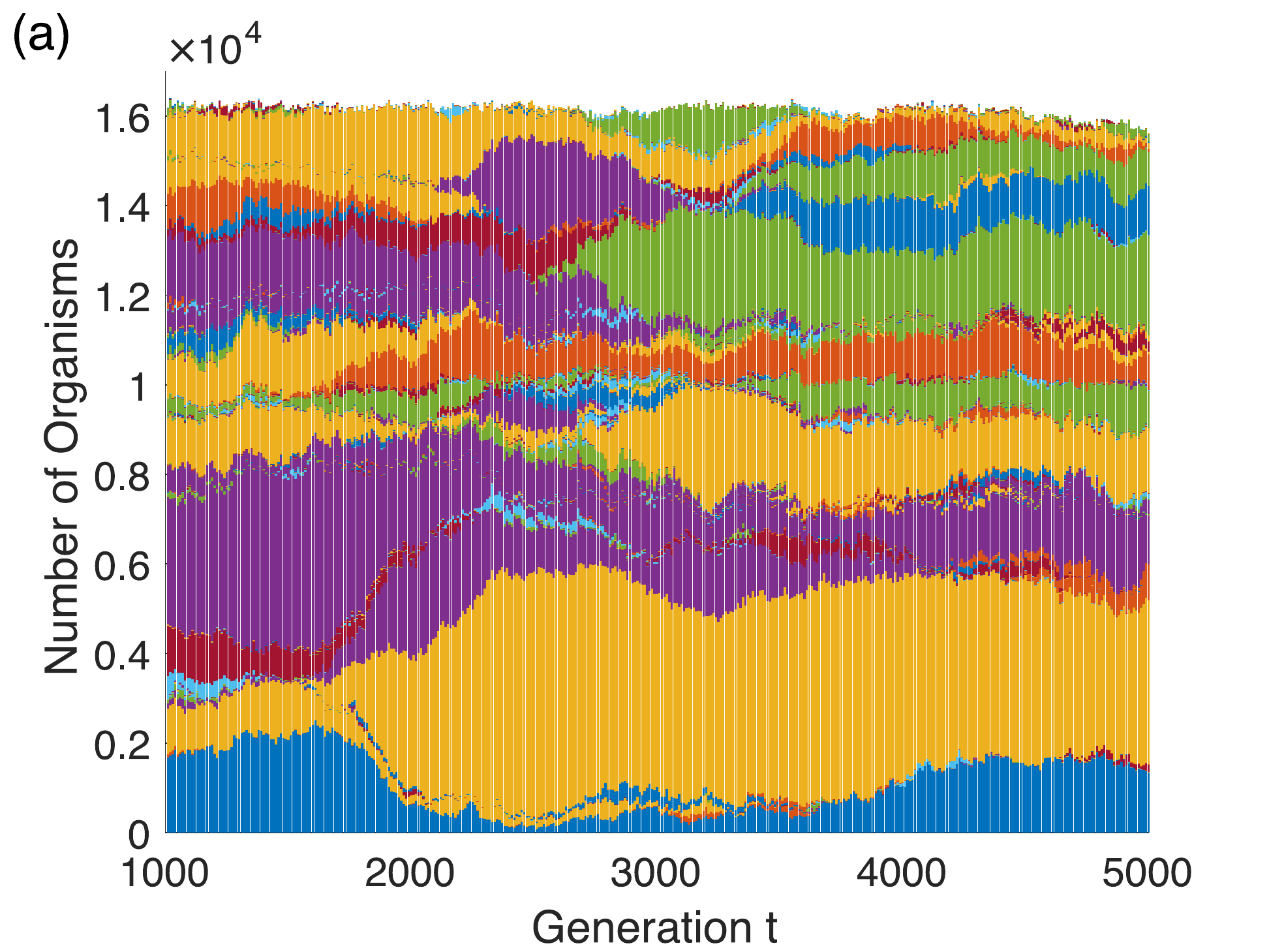}}\hspace*{1em}
  \subfigure { \label{fig:16metabo2}
    \includegraphics[width = 0.48\linewidth]{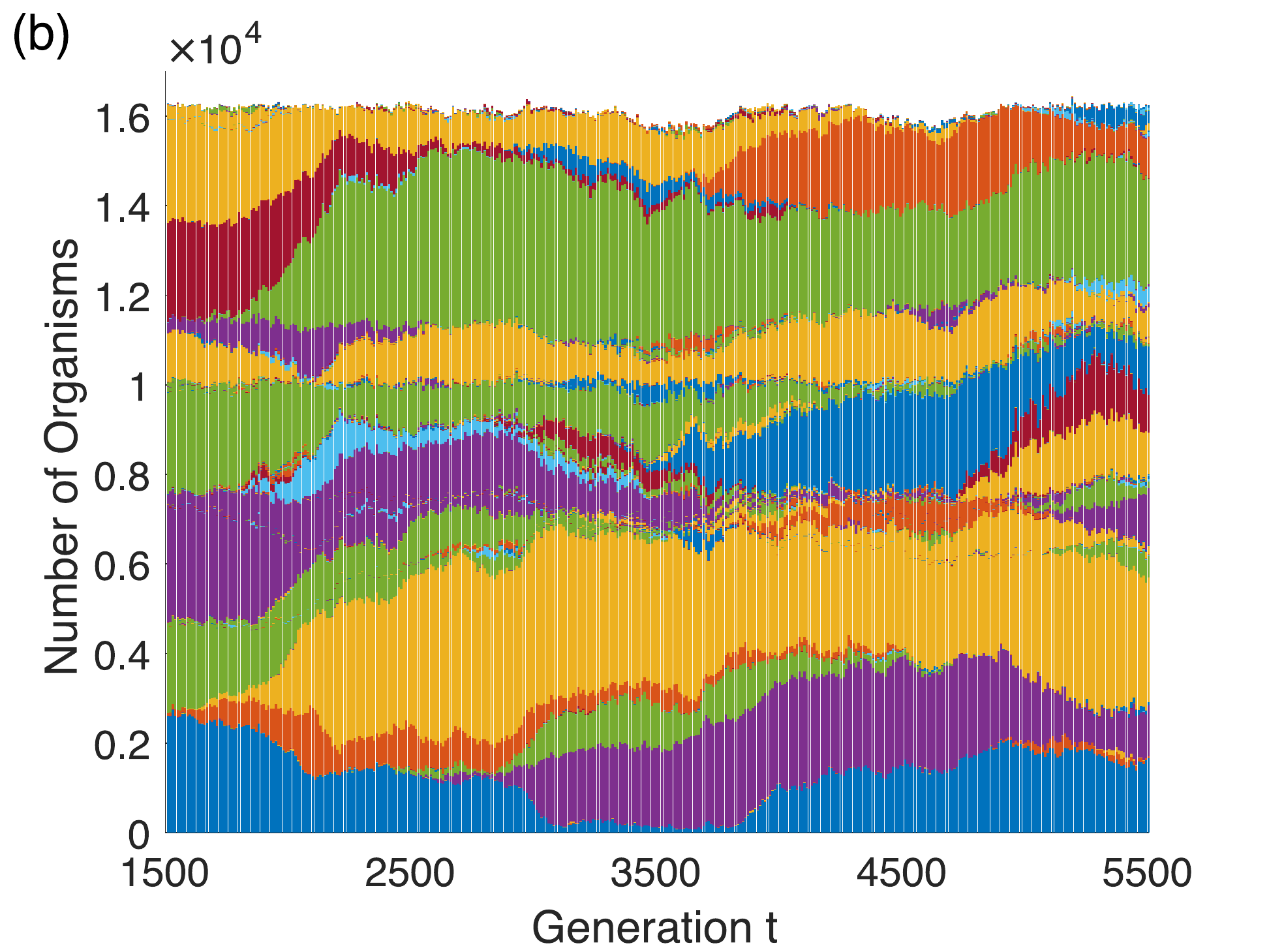}}
  \subfigure { \label{fig:16metabo3}
    \includegraphics[width = 0.48\linewidth]{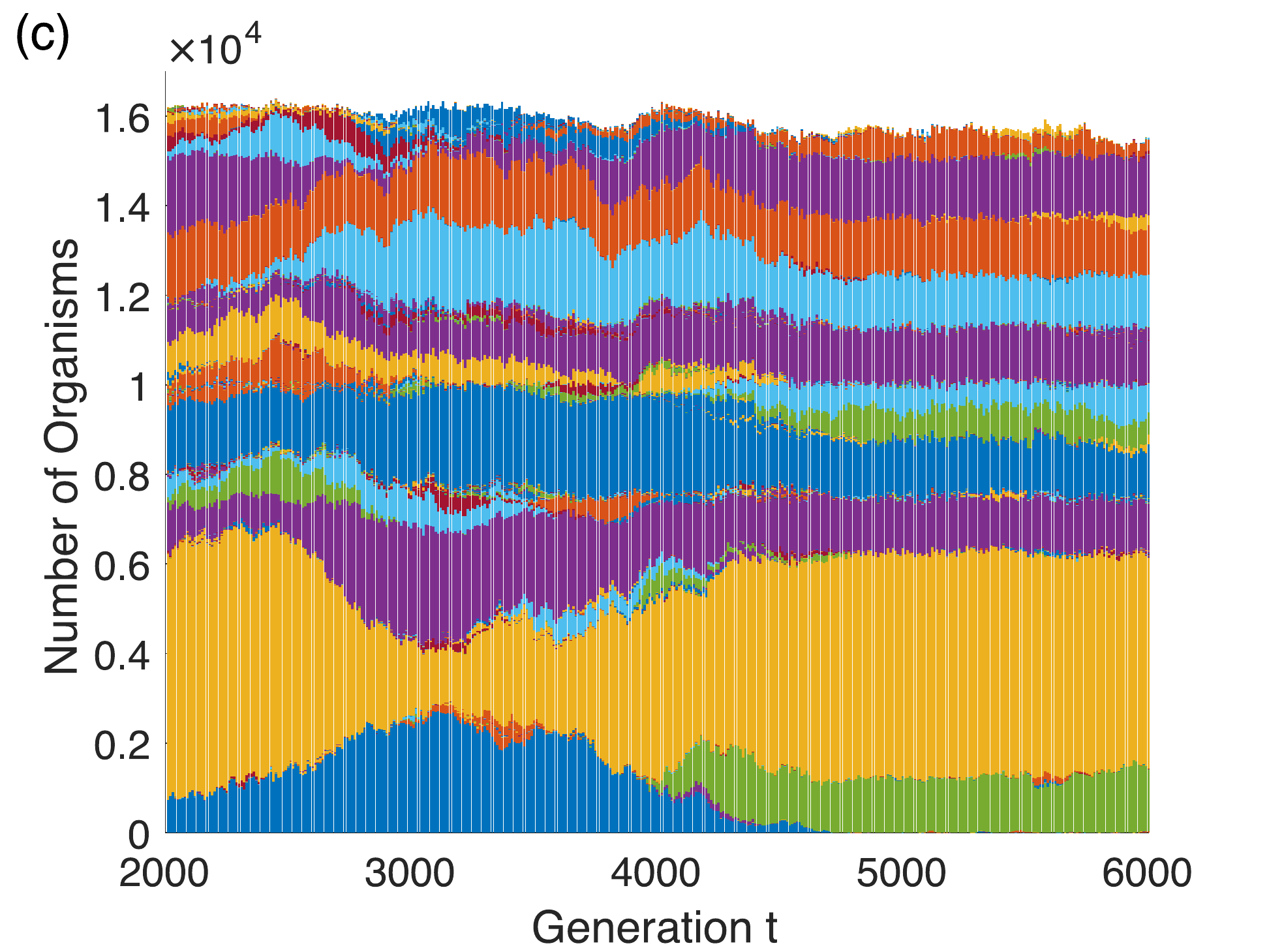}}\hspace*{1em}
  \subfigure {
    \includegraphics[width = 0.48\linewidth]{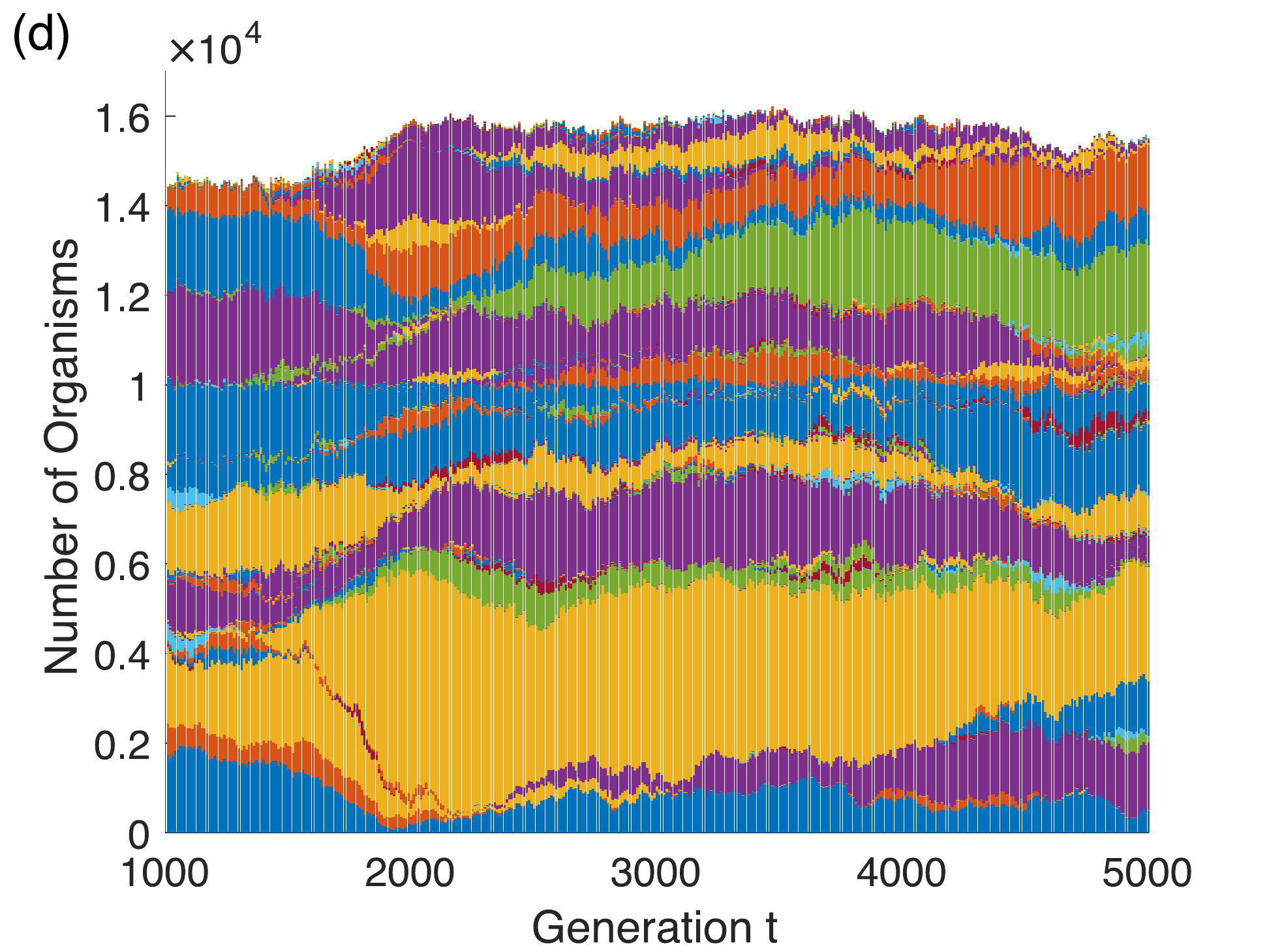}}
\caption{The changes of total population divided into species type over $4000$ generations, for four different runs but with the same parameters and initial conditions, in the 16-metabolite system with inflowing metabolite $\overline{10}$. Since there are too many species, some species are denoted by the same colour, but two different species with the same colour are always gapped. One generation in simulation basically means every organism in the system completes one life cycle.}
\label{fig:16metabo}
\end{figure*}

Figure \ref{fig:16metabo} shows another example, four different runs in the 16-metabolite system with inflowing metabolite $\overline{10}$ (with the same parameters and initial conditions). Instead of visualising specific demographic structures, we show the changes of total population divided into species type, because there are in total $n(n+1)/2 = 136$ possible species, which makes it hard to see the interactions by the previous visualization.
We can observe that (1) In each run, many different demographic structures appear, e.g., that at $t = 1500$ and $4000$ in Fig. \ref{fig:16metabo1} and that at $t = 3500$ and $5000$ in Fig. \ref{fig:16metabo2}, not only within each figure but also between these figures; (2) Some of these demographic structures can last for quite a long time, e.g., that from $t = 4500$ to $6000$ in Fig. \ref{fig:16metabo3}, while some are transient, e.g., that at $t = 2500$ in Fig. \ref{fig:16metabo1}; (3) No single equilibrium is always stable, which is a general observation in the number soup model. These facts indicate rich biodiversity.

\begin{figure*}[tbhp] \centering
  \subfigure {
    \includegraphics[width = 0.48\linewidth]{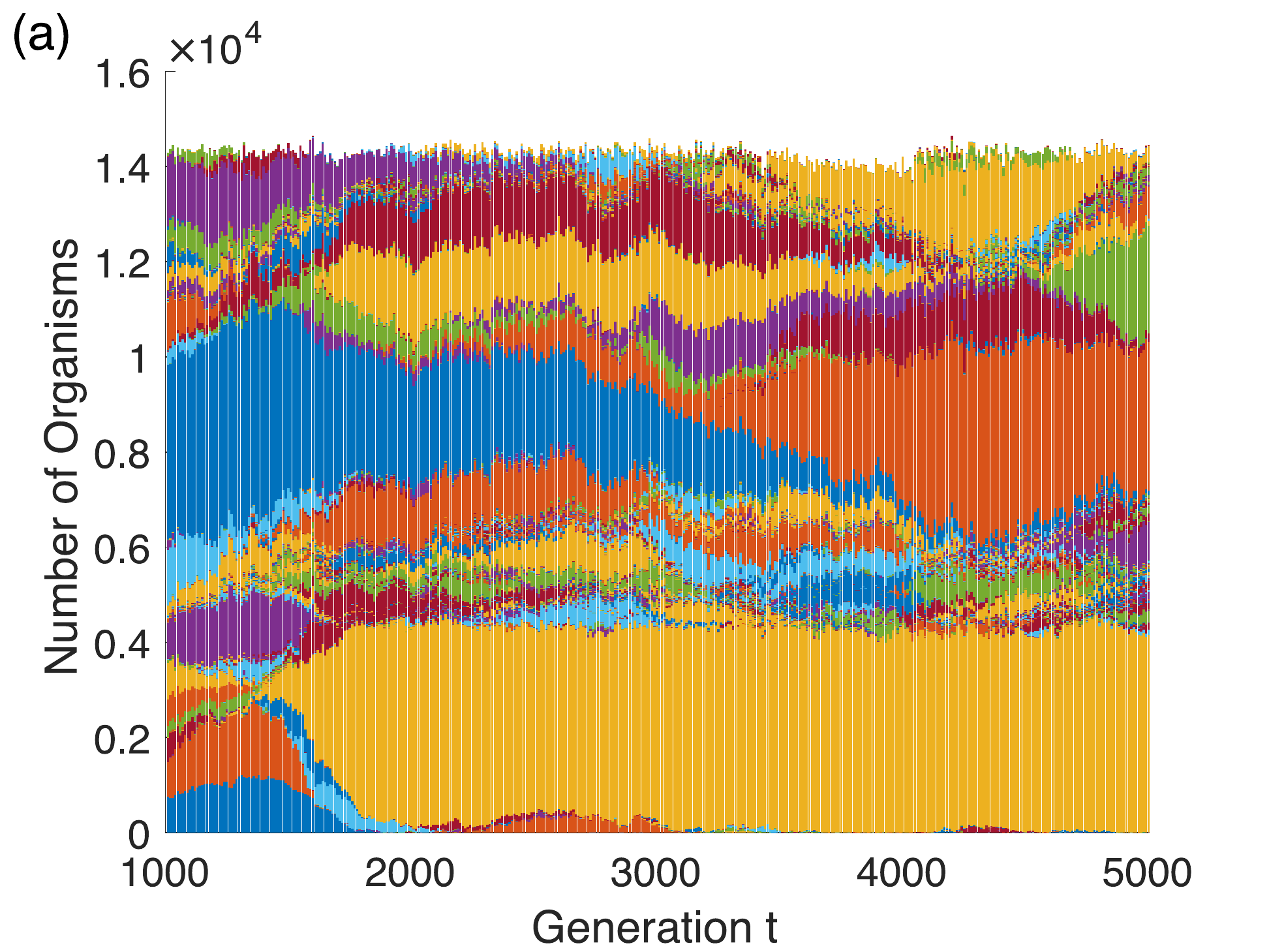}}\hspace*{1em}
  \subfigure {
    \includegraphics[width = 0.48\linewidth]{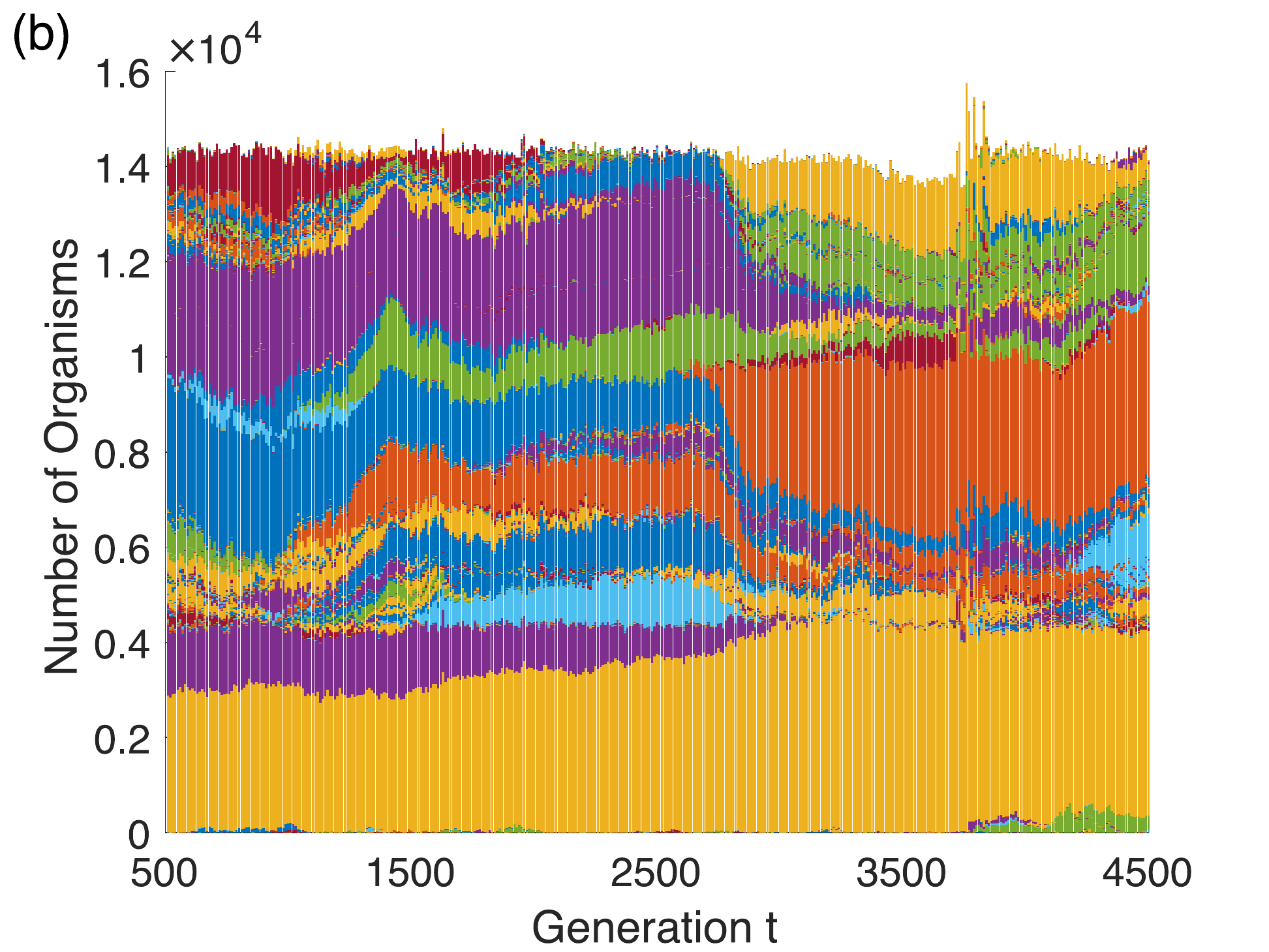}}
  \subfigure {
    \includegraphics[width = 0.48\linewidth]{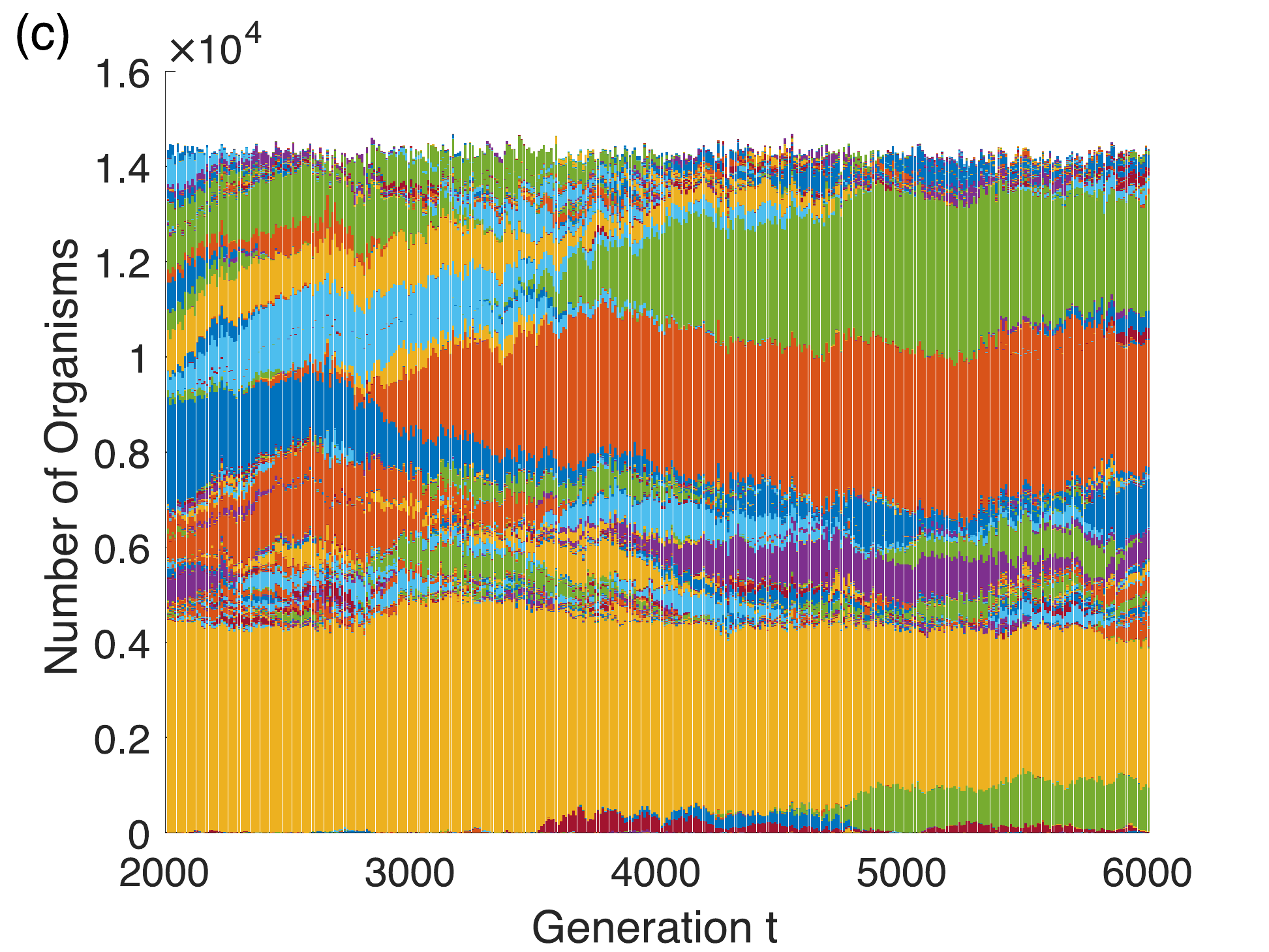}}\hspace*{1em}
  \subfigure { \label{fig:39metabo4}
    \includegraphics[width = 0.48\linewidth]{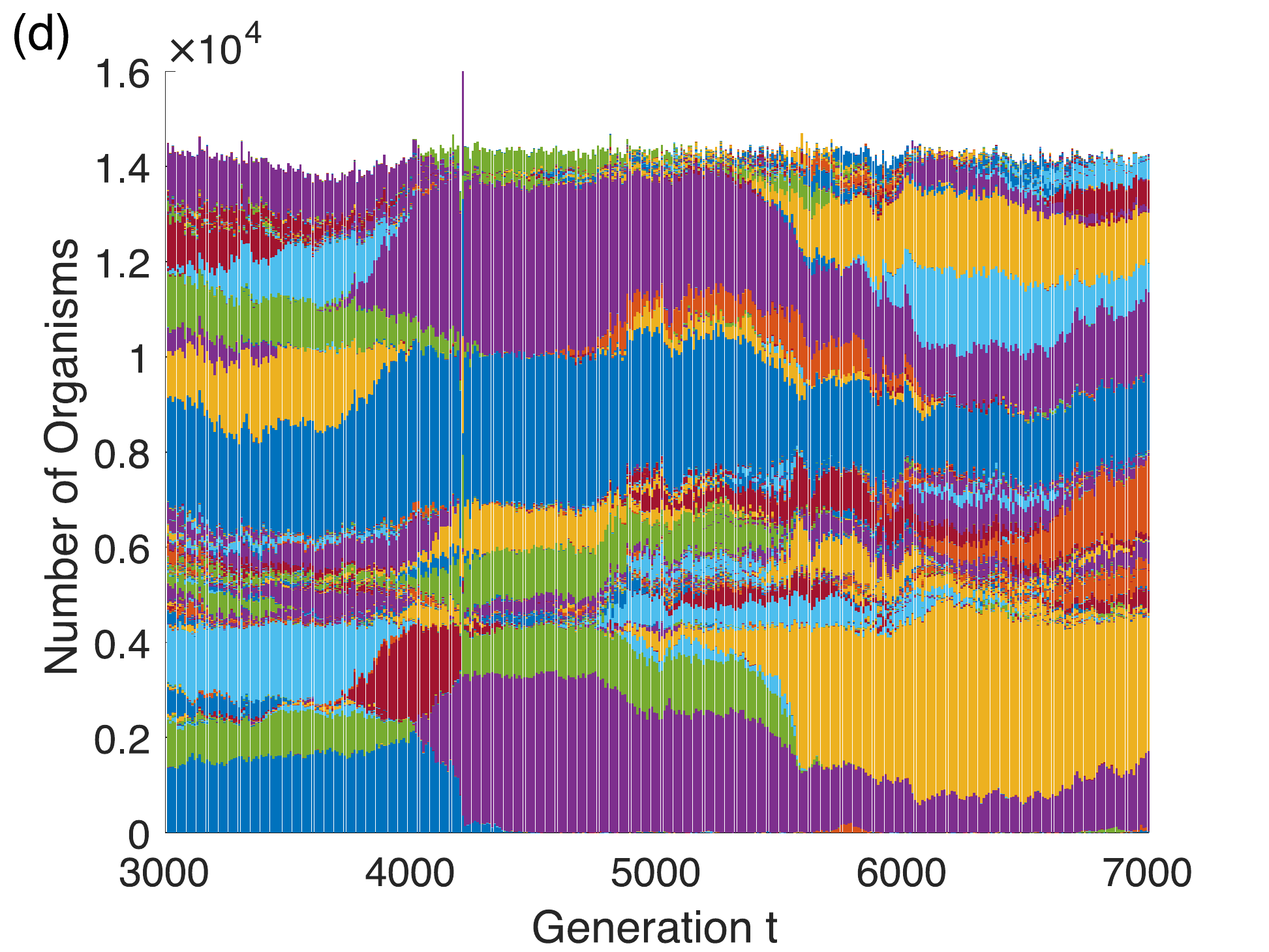}}
\caption{The changes of total population divided into species type over $4000$ generations, for four different runs but with the same parameters and initial conditions, in the 39-metabolite system with inflowing metabolite $\overline{17}$.}
\label{fig:39metabo}
\end{figure*}

Figure \ref{fig:39metabo} shows a more complex example, in the 39-metabolite system with inflowing metabolite $\overline{17}$, where there are in total $780$ possible species.
We observe there are much more different demographic structures ever appeared.

Another observation from Fig. \ref{fig:39metabo} is that the yellow species (at the bottom in each figure) is always very abundant. Note that the yellow species is $s_{1, 17}$ and the inflowing metabolite is $\overline{17}$ (The reason why $s_{17, 17}$ is not more abundant than $s_{1, 17}$ is that $s_{1, 17}$ is more efficient in obtaining $\overline{17}$, referring to the paper \cite{Liu17IIR} in Section 3.1). So there may exist some common features for most of the possible demographic structures, but it does not necessarily mean that some species would dominate in every demographic structure, e.g., there is no $s_{1, 17}$ in the demographic structure in Fig. \ref{fig:39metabo4} from $t = 3000$ to $5500$. The same situation occurs in the last example in Fig. \ref{fig:16metabo}, that the yellow species (at the bottom in each figure), which is $s_{1, 10}$, is also very abundant in each figure.

Moreover, in both Fig. \ref{fig:16metabo} and \ref{fig:39metabo}, the total population size is almost constant all the time, which depends on the inflow rate. It seems a general property. The explanation of this property may be hidden in the following two facts: (1) The whole ecosystem would always self-organise itself to fully consume the inflowing metabolites, as discussed in the paper \cite{Liu17IIR} Section 3.1, so the rate of consumption of the inflow, namely the rate of metabolites escaping from the system, is fixed (since the inflow rate is fixed); (2) The mathematical property of modular addition makes the numbers change in a certain way, which makes the total population fixed, referring to the paper \cite{Liu17IIR} supplementary material Section (g) about the discussion of the number of species loops in equilibrium. This observation gives some insights into the question whether an ecosystem has a maximum biomass production rate for a fixed inflow rate.

\section{Properties of species loop}
\label{sec:speciesloop}

We defined a useful ecological concept for looking at inter-dependencies in ecosystems: the species loop. Our definition arises from the term ``microbial loop'', that acts as a sink of carbon in a water column \cite{Fenchel2008}. It is an influential concept in biological oceanography \cite{Azam1983}. In our artificial ecosystem, at any one time various species loops persist, self-organising to consume up all the available resources. That is, any demographic structure which is able to persist in the ecosystem can be considered as being constituted by different species loops. This is a general property for any $n$-metabolite system with any inflow.

We have introduced the scheme to calculate the compositions of any demographic structure, referring to the paper \cite{Liu17IIR} supplementary material Section (h). Here are more examples of different demographic structures. In the 9-metabolite system with inflowing metabolite $\bar{7}$, the demographic structure shown in Fig. \ref{fig:9metabo1} consists of
\begin{equation*} \begin{cases} \begin{split}
  40.0\% & \text{~~of~~the~~species~~loop} \\
  	& 5s_{11} + 2s_{22} + s_{25} + s_{44} + 3s_{47} + 3s_{77} + s_{78}~~(\Omega = 9) \\
  33.3\% & \text{~~of~~the~~species~~loop} \\
   	& 4s_{11} + s_{22} + s_{25} + s_{26} + 2s_{44} + s_{47} \\
	& \qquad \qquad \qquad \qquad \qquad~ + 4s_{77} + s_{78} + s_{88}~~(\Omega = 9) \\
   26.7\% & \text{~~of~~the~~species~~loop} \\
   	& 6s_{11} + 3s_{22} + 5s_{47} + 2s_{77}~~(\Omega = 9)
\end{split} \end{cases} \end{equation*}
The demographic structure shown in Fig. \ref{fig:9metabo2} consists of
\begin{equation*} \begin{cases} \begin{split}
  50.0\% & \text{~~of~~the~~species~~loop} \\
  	& s_{11} + 4s_{17} + s_{18} + s_{23} + 2s_{55} + s_{67} \\
	& \qquad \qquad \qquad \qquad \qquad~ + 3s_{78} + s_{79}~~(\Omega = 9) \\
  34.6\% & \text{~~of~~the~~species~~loop} \\
   	& s_{11} + 3s_{17} + 3s_{18} + s_{23} + s_{36} + s_{47} \\
	& \qquad \qquad \qquad ~~~~ + s_{59} + 2s_{67} + 3s_{79}~~(\Omega = 9) \\
   11.5\% & \text{~~of~~the~~species~~loop} \\
   	& 2s_{11} + 3s_{17} + s_{18} + s_{22} + s_{36} + s_{47} + s_{55} \\
	& \qquad \qquad \qquad \quad~ + s_{67} + 2s_{78} + 2s_{79}~~(\Omega = 9) \\
   3.9\% & \text{~~of~~the~~species~~loop} \\
   	& 2s_{11} + 4s_{17} + s_{22} + s_{47} + 2s_{55} + 4s_{78}~~(\Omega = 9)
\end{split} \end{cases} \end{equation*}
The demographic structure shown in Fig. \ref{fig:9metabo3} consists of
\begin{equation*} \begin{cases} \begin{split}
  40.0\% & \text{~~of~~the~~species~~loop} \\
  	& 3s_{11} + 8s_{17} + s_{25} + 2s_{27} + s_{33} + 3s_{36} \\
	& \qquad \qquad \qquad + 5s_{67} + s_{69} +4s_{79} + 4s_{88}~~(\Omega = 18)\\
  20.0\% & \text{~~of~~the~~species~~loop} \\
   	& 7s_{17} + 2s_{36} + s_{49} + s_{67} + s_{68} + s_{79} + 3s_{88}~~(\Omega = 9) \\
   20.0\% & \text{~~of~~the~~species~~loop} \\
   	& 14s_{17} + 3s_{27} + 3s_{48} + s_{55} + 3s_{68} + 2s_{69} \\
	& \qquad \qquad \qquad \qquad \qquad \qquad + s_{79} + 4s_{88}~~(\Omega = 18) \\
   20.0\% & \text{~~of~~the~~species~~loop} \\
   	& 3s_{11} + 8s_{17} + 2s_{23} + s_{27} + s_{33} + 2s_{36} \\
	& \qquad \qquad \qquad + s_{55} + 6s_{67} + 3s_{79} + 4s_{88}~~(\Omega = 18)
\end{split} \end{cases} \end{equation*}
The demographic structure shown in Fig. \ref{fig:9metabo4} consists of
\begin{equation*} \begin{cases} \begin{split}
  54.3\% & \text{~~of~~the~~species~~loop} \\
  	& s_{13} + 3s_{14} + 3s_{15} + 2s_{37} + 3s_{67} + 2s_{77}~~(\Omega = 9)\\
  45.7\% & \text{~~of~~the~~species~~loop} \\
   	& 2s_{13} + 3s_{14} + 2s_{15} + 2s_{45} + 2s_{67} + 3s_{77} \\
	& \qquad \qquad \qquad \qquad \qquad ~~~~ + s_{78} + s_{99}~~(\Omega = 9) 
\end{split} \end{cases} \end{equation*}

There are three properties of species loop that should be noted and are worthy of further investigation in future works.

First, the number of types of coexisting species loops has a limitation, which is associated with the number of types of metabolites and species involved in the system. For a particular demographic structure, recall the matrix $\mathbf{Z}$ (referring to the paper \cite{Liu17IIR} supplementary material Section (h) step vii), of which the columns are all the ``quasi'' species loops possible to constitute the reduced population vector of this demographic structure. So the number of types of coexisting species loops in this demographic structure is at most the number of independent columns of $\mathbf{Z}$, namely rank($\mathbf{Z}$). On the other hand, rank($\mathbf{Z}$) is equal to the number of free parameters of $\mathbf{x}$, consequently equal to the number of variables of $\mathbf{x}$ minus rank($\mathbf{A}$), consequently equal to the number of columns of $\mathbf{A}$ minus rank($\mathbf{A}$). Note that the number of columns of $\mathbf{A}$ is the number of species involved in this demographic structure (denoted by $K$). And also note that $\mathbf{A}$ is a matrix with $W - 1$ rows, so always rank($\mathbf{A}$) $\leqslant W - 1$, where $W$ is the number of types of metabolites involved and ``$1$'' is the number of types of inflowing metabolites, later denoted by $I$. In the number soup model, the equality always holds, namely rank($\mathbf{A}$) $= W - 1$, and generally rank($\mathbf{A}$) $= W - I$.

Therefore, in the number soup model, the number of types of coexisting species loops is at most $K - W + I$. The insight into living ecosystem is that (1) the number of coexisting species loops is constrained (although we do not know the constraints for individual species), so biodiversity should also be constrained; (2) the number of types of metabolites involved, namely the intermediate products, provides the constraints; and (3) species loops could be considered as a higher level unit of natural selection than individual species (since the constrains acts on species loops), as a quasi-unit of community ecology, which natural selection indirectly acts on.

The second property of species loop is that for a relatively complex demographic structure, it can be interpreted as being constituted by different sets of species loops; while for a relatively simple demographic structure, only one interpretation is possible. For example, for the three relatively simple demographic structures shown in Fig. \ref{fig:9metabo1}, \ref{fig:9metabo2} and \ref{fig:9metabo4}, the compositions given above are the only possibilities. However, for the complex demographic structure shown in Fig. \ref{fig:9metabo3}, there are more interpretations. Besides the one above, it can also be interpreted as being constituted by
\begin{equation*} \begin{cases} \begin{split}
  40.0\% & \text{~~of~~the~~species~~loop} \\
  	& 3s_{11} + 8s_{17} + s_{25} + 2s_{27} + s_{33} + 3s_{36} \\
	& \qquad \qquad~~~ + 5s_{67} + s_{69} +4s_{79} + 4s_{88}~~(\Omega = 18)\\
  20.0\% & \text{~~of~~the~~species~~loop} \\
   	& 7s_{17} + s_{27} + s_{36} + s_{48} + s_{49} + 2s_{68} \\
	& \qquad \qquad \qquad \qquad \qquad \quad~~ + s_{79} + 2s_{88} ~~(\Omega = 9) \\
   20.0\% & \text{~~of~~the~~species~~loop} \\
   	& s_{11} + 12s_{17} + 3s_{27} + s_{36} + 2s_{48} + s_{55} + s_{67} \\
	& \qquad \qquad ~~~ + 2s_{68} + 2s_{69} + 2s_{79} + 4s_{88}~~(\Omega = 18)\\
   20.0\% & \text{~~of~~the~~species~~loop} \\
   	& 2s_{11} + 10s_{17} + 2s_{23} + s_{33} + 2s_{36} + s_{55} \\
	& \qquad \qquad \qquad \qquad + 6s_{67} + 2s_{79} + 5s_{88}~~(\Omega = 18)
\end{split} \end{cases} \end{equation*}
or constituted by
\begin{equation*} \begin{cases} \begin{split}
  40.0\% & \text{~~of~~the~~species~~loop} \\
  	& s_{11} + 12s_{17} + s_{25} + s_{33} + 3s_{36} + 5s_{67} \\
	& \qquad \qquad \qquad \qquad~~ + s_{69} + 2s_{79} + 6s_{88}~~(\Omega = 18)\\
  20.0\% & \text{~~of~~the~~species~~loop} \\
   	& 7s_{17} + s_{27} + s_{36} + s_{48} + s_{49} + 2s_{68} \\
	& \qquad \qquad \qquad \qquad \qquad \qquad + s_{79} + 2s_{88}~~(\Omega = 9) \\
   20.0\% & \text{~~of~~the~~species~~loop} \\
   	& 5s_{11} + 4s_{17} + 7s_{27} + s_{36} + 2s_{48} + s_{55} \\
	& \qquad \qquad ~~~~~ + s_{67} + 2s_{68} + 2s_{69} + 6s_{79}~~(\Omega = 18)\\
   20.0\% & \text{~~of~~the~~species~~loop} \\
   	& 2s_{11} + 10s_{17} + 2s_{23} + s_{33} + 2s_{36} + s_{55} \\
	& \qquad \qquad \qquad \qquad~ + 6s_{67} + 2s_{79} + 5s_{88}~~(\Omega = 18)
\end{split} \end{cases} \end{equation*}
Since there are different interpretations for the same demographic structure, we thus say the species loop is a quasi-unit of natural selection, rather than a real unit.

\begin{figure}[h] \centering
    \includegraphics[width = 0.5\linewidth]{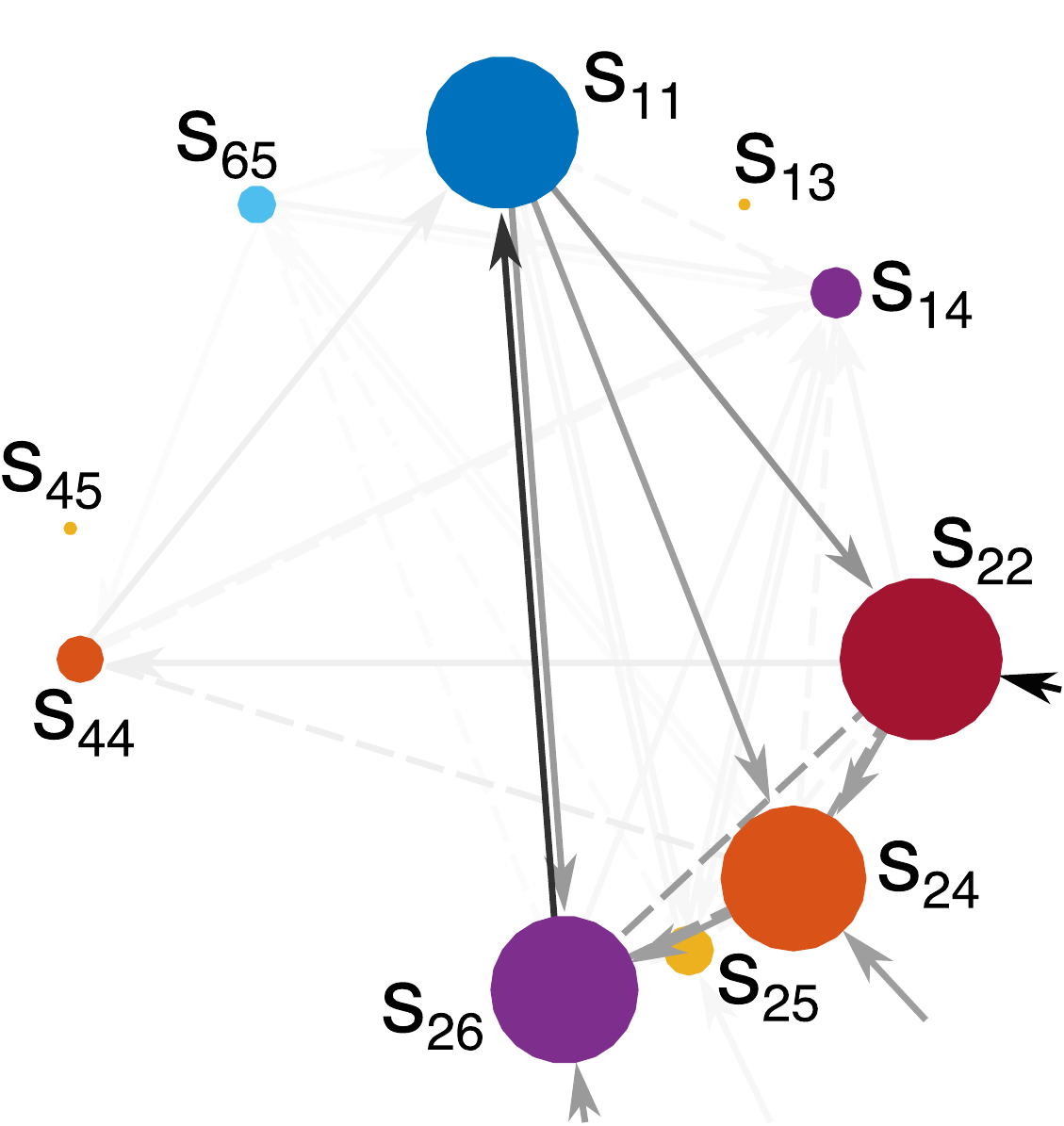}
\caption{The demographic structure arising in an evolution in the system $n = 6$ and $u = 2$, adapted from the paper \cite{Liu17IIR}. It consists of the species loop $s_{14} + s_{22} + s_{25}$ ($\Omega = 3$), the species loop $s_{11} + 2s_{22} + s_{44} ~(\Omega = 3)$, and the species loop $s_{11} + s_{22} + s_{24} + s_{26} ~(\Omega = 3)$.}
\label{fig:6meta}
\end{figure}

The third property is about $\Omega$, the number of species loops constituting one demographic structure. For example, all of the three species loops that constitute the demographic structure shown in Fig. \ref{fig:6meta} (in the 6-metabolite system) have $\Omega = n/2$, referring to the paper \cite{Liu17IIR} supplementary material Section (h).
For another example, all of the species loops that constitute the demographic structure shown in Fig. \ref{fig:9metabo1} (in 9-metabolite system) have $\Omega = n$; in Fig. \ref{fig:9metabo2} all of them have $\Omega = n$; in Fig. \ref{fig:9metabo3} three of them have $\Omega = 2n$ and one has $\Omega = n$; and in Fig. \ref{fig:9metabo4} all of them have $\Omega = n$. The question raised is that: In an $n$-metabolite system, do $\Omega$ and $n$ always have common factors?

\section{Intuition of Gillespie Algorithm}
\label{sec:Gill}

If the process that organisms take actions one by one is a Poisson process, the number of events---an event means that an organism take an action---in time interval $(t, t+\tau]$ then follows a Poisson distribution with parameter $\lambda \tau$, i.e.,
\begin{equation}
	P(N(t+\tau)-N(t)=k)=\frac{e^{-\lambda\tau}(\lambda\tau)^k}{k!},\qquad k=0,1,\dots
\label{eq:A1}
\end{equation}
where $k$ is the number of events. We can prove that the waiting time from the current event to the next event follows an exponential distribution with parameter $\lambda$, as follows.

\eqref{eq:A1} with $k=0$ tells us the probability that no event occurs during $\tau$, i.e.,
\begin{equation*} \begin{split}
    P(\text{no event occurs during $\tau$}) & =P(N(t+\tau)-N(t)=0) \\
   & =\frac{e^{-\lambda\tau}(\lambda\tau)^0}{0!}=e^{-\lambda\tau}
\end{split} \end{equation*}
So, the probability that next event occurs after $\tau$ is
\begin{equation} \begin{split}
    & P(\text{next event occurs after $\tau$}) \\
    = & P(\text{no event occurs during $\tau$}) = e^{-\lambda\tau}
\label{eq:A2}
\end{split} \end{equation}

On the other hand, if the waiting time till next event follows an exponential distribution, we can write it as
\begin{equation}
	P(w,w+dw)=\lambda e^{-\lambda w}dw
\label{eq:A3}
\end{equation}
where $w$ is the time of the current event. So the probability that the next event occurs after a given time $\tau'$ is
\begin{equation} \begin{split}
    P(\text{next event occurs after $\tau'$}) = & \int_{\tau'}^{\infty}\lambda e^{-\lambda w}dw \\
    = & [-e^{-\lambda w}]_{\tau'}^{\infty} = e^{-\lambda \tau'}
\label{eq:A4}
\end{split} \end{equation}

Note that \eqref{eq:A2} and \eqref{eq:A4} have exactly the same format. Therefore, if a process is a Poisson process (i.e., follows \eqref{eq:A1}), the waiting time till next event occurs follows an exponential distribution, i.e., follows \eqref{eq:A3}. Proof ends.

Now we assume that the process those organisms take actions  one by one is a Poisson process, and then we can decide $\Delta t$, the waiting time till next event occurs, by choosing a random number from an exponential distribution with parameter $\lambda$.

But, what is $\lambda$? Here we set it as the total number of organisms, $N$. Intuitively, we expect that one unit time is passed if every organism takes action once. That is, the average time interval between two actions is $1 / N$. On the other hand, for the exponential distribution with $\lambda$, the expected value is $1 / \lambda$. So if we set $\lambda = N$, the average waiting time is $1/ N$, which accords with the intuition.

Note that, $N$ itself is also a function of time. The scheme described above to decide the time intervals between two events is the standard scheme of Gillespie Algorithm.

\section{Growth rate comparing with Monod equation}
\label{sec:Monod}

The Monod equation is a classic empirical law for microbial growth, written as
\begin{equation*}
    \frac{1}{S}\frac{dS}{dt}\equiv\mu=\mu_{\text{max}}\frac{R}{K+R}
\end{equation*}
where $S$ is the population or biomass of microbes, $\mu$ is the specific growth rate, $\mu_{\text{max}}$ is the maximum specific growth rate, $R$ is the concentration of the limiting substrate for growth, and $K$ is the ``half-velocity constant'', namely the value of $R$ when $\mu/\mu_{\text{max}}=0.5$ (both $\mu_{\text{max}}$ and $K$ are empirical coefficients). If there is more than one limiting substrate, the right side should be multiplied by another term $R'/(K'+R')$.

Comparing with our number soup model \cite{Liu17IIR}, $\mu_{\text{max}}$ should be $1$ and $K$ should be $a$. So if we employ the Monod equation, the growth equation for species $s_{12}$ should be written as (the same reasoning for other species)
\begin{equation*}
    \frac{1}{S_{12}}\frac{dS_{12}}{dt}=\frac{R_1}{a+R_1}\frac{R_2}{a+R_2}\equiv q(R_1, R_2)
\end{equation*}
But in our model, the equation is (referring to the differential equations in the supplementary material of the paper \cite{Liu17IIR})
\begin{equation} \label{eq:Monod2}
\begin{split}
    \frac{1}{S_{12}}\frac{dS_{12}}{dt} & =2\frac{R_1}{a+R_1}\frac{R_2}{a+R_2}-1 \\
    & \equiv q(R_1, R_2)-(1-q(R_1, R_2))
\end{split} \end{equation}
The common term is $q(R_1, R_2)$---the probability of being able to obtain metabolites---represents the birth process. The term $-(1-q(R_1, R_2))$ only appears in \eqref{eq:Monod2}, which is the probability of being unable to obtain enough metabolites. It thus represents the death process. Therefore, in our model, there are both birth and death processes, while in the Monod scheme, there is only birth process.

Note that one unit time in our model is basically the doubling time of those organisms. So in short-term experiments, e.g., for a few doubling time, the Monod equation makes sense since organisms may survive for a few doubling times even without taking food. However, in long-term experiments, starvation has to be taken into account. In addition, we are only interested in living organisms, so dead organisms need to be excluded. Overall, our scheme makes more sense.

\section{Another way to think what a species excretes}
\label{sec:addition}

In the $n$-metabolite system, what species $s_{ij}$ excretes is based on modulus-$(n+1)$ addition (Eq. (2.2) in the paper \cite{Liu17IIR}), namely,
\begin{equation*} \begin{aligned}
s_{ij}: ~~\bar{i} + \bar{j} \rightarrow 
  \begin{cases}
    \overline{i+j}, & \text{~if~} i + j < n + 1 \\
    \bar{1},   & \text{~if~} i + j = n + 1\\
    \bar{1} + \overline{(i+j) \text{~mod~} (n+1)}, & \text{~if~} i + j > n + 1
  \end{cases}
\end{aligned} \end{equation*}
It is a bit different from the normal modular addition on the set $\{ \bar{0}, \bar{1}, \bar{2}, \cdots, \bar{n} \}$, because by using the equations above, we have put an additional $\bar{1}$ if the addition produces a ``wrap around'' number, and always got rid of $\bar{0}$.

If it feels unnatural, there is another way to think about it, which results in identical equations as above: In the $n$-metabolite system, what species $s_{ij}$ excretes is based on the addition in base-$(n+1)$ numeral system with the settings that (1) the resulted double-digit number splits into the two digits constituting itself and (2) the number zero always disappears. For example, in the $3$-metabolite system, species $s_{33}$ excretes $\bar{1}$ and $\bar{2}$ because $3+3 = 12$ in base-$4$ numeral system and $12$ splits into $1$ and $2$; species $s_{22}$ excretes $\bar{1}$ because $2+2 = 10$ in base-$4$ numeral system and $10$ splits into $1$ and $0$, and then $0$ disappears; species $s_{12}$ excretes $\bar{3}$ because $1+2 = 3$ in base-$4$ numeral system.

\section{Types of cross-feeding}
\label{sec:typesCF}

Cross-feeding occurs when a microbe utilises metabolites excreted by other microbes \cite{Hummert14EGT}. In terms of whether excretion is costly, \textit{incidental} and \textit{cooperative} cross-feeding can be distinguished \cite{Estrela12FMT}. The former one refers to that the excretion is a waste product and thus has no cost to the producer, while the cooperative cross-feeding refers to that the excretion is costly to the producer but the up-front investment may be paid back by the partner species using the excretion.

From another perspective, \textit{sequential} and \textit{reciprocal} cross-feeding can be distinguished. The former refers to that a microbe feeds on the intermediates or waste products of the other, but does not offer any benefit to the feeder, while the reciprocal cross-feeding refers to the mutually beneficially exchange of metabolites \cite{Hummert14EGT, Estrela10EOC}.

Therefore, the number soup model only talks about the incidental cross-feeding, without involving the cooperative cross-feeding. From the other perspective, the model does not particularly distinguish the sequential and the reciprocal cross-feeding, so both of them can appear.

\section{Reflections on cooperators and cheaters}
\label{sec:cheater}

There is a big question in ecology why cooperators and cheaters could coexist since cheaters always get benefits but do not pay anything back. Some authors claim that spatial effects would promote and maintain coexistence \cite{Rainey98ARI, Pfeiffer03AES}. However, is it possible for coexistence if the environment is homogeneous or well-mixed?

According to the number soup model, besides other non-spatial models, the answer is yes. But note that our model only deals with the incidental cross-feeding, which means that cooperation carries no cost. So the meanings of ``cooperators'' and ``cheaters'' in this section are a bit narrower than the normal definitions. It is more like what Connor called as ``pseudo-reciprocity'' \cite{Connor86PII}. It has been suggested that in many mutualism scenarios, one of the partners may be largely passive and bear no costs from the interaction \cite{Foster06AGM}.

\begin{figure*}[tbp] \centering
  \subfigure{\label{fig:case3gra_1}
    \includegraphics[width=0.23\linewidth] {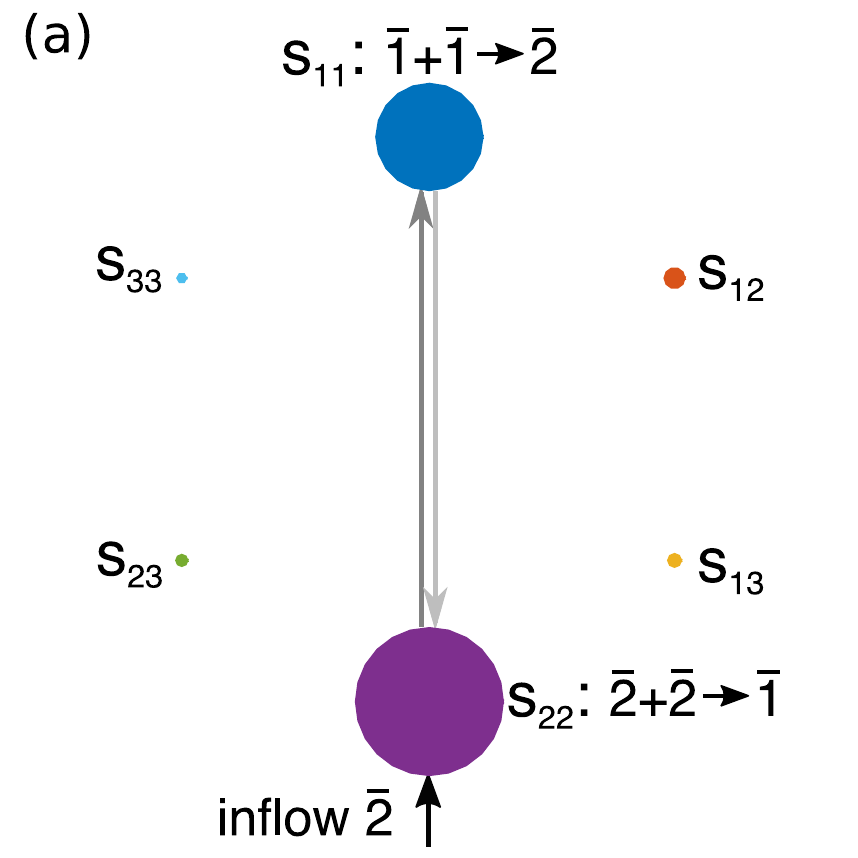}}
  \subfigure{\label{fig:case3gra_inter}
    \includegraphics[width=0.23\linewidth] {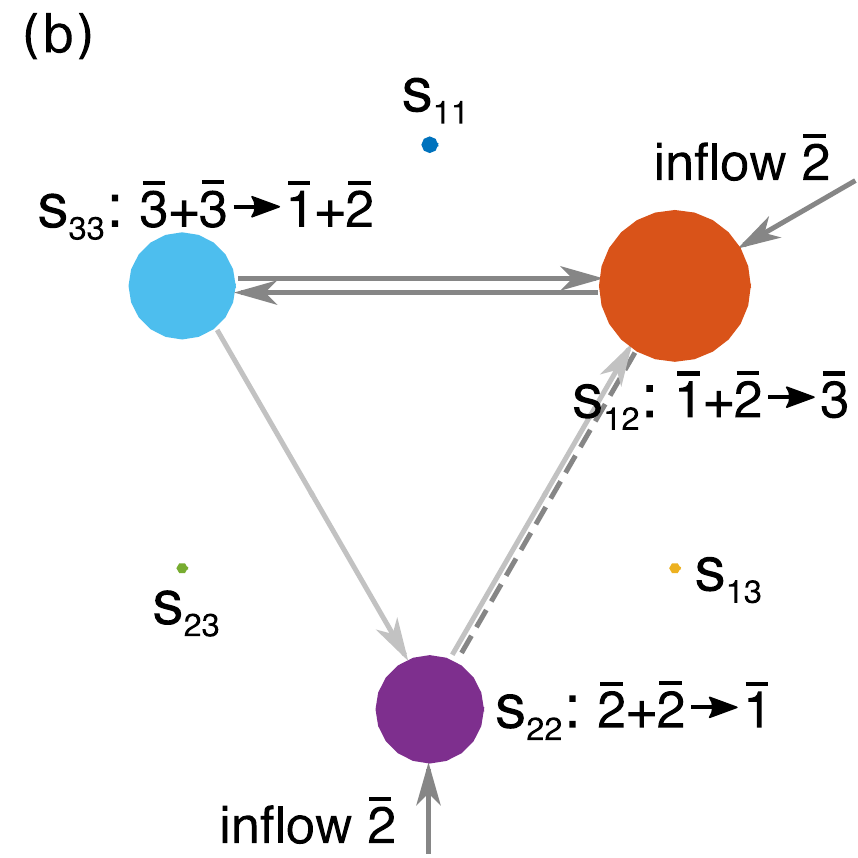}}
  \subfigure{\label{fig:case3gra_3}
    \includegraphics[width=0.23\linewidth] {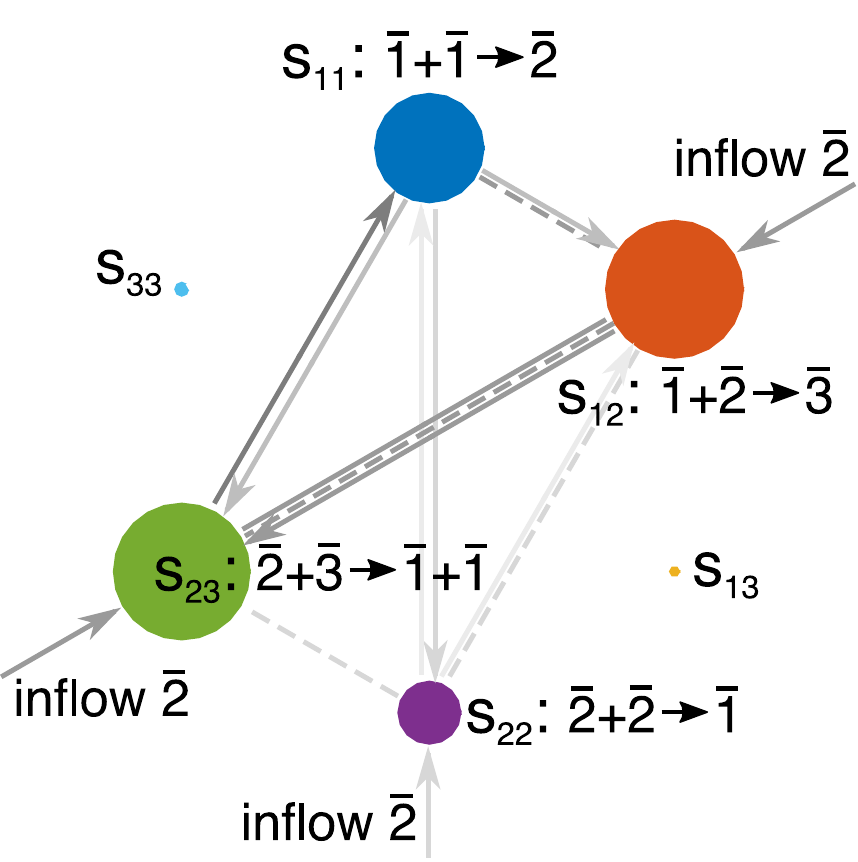}}
  \subfigure{\label{fig:case3gra_2}
    \includegraphics[width=0.23\linewidth] {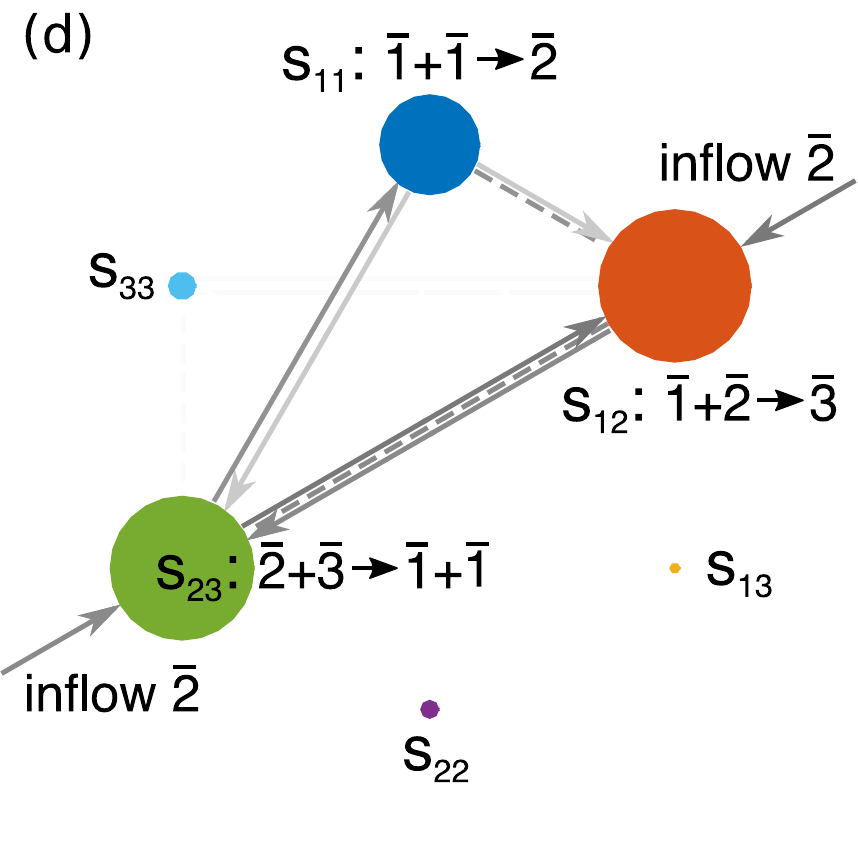}}
  \caption{Demographic structures arising in an example evolution in the system with $n = 3$ and $u = 2$. (a) The demographic structure occurred at generation $t = 400$. (b) The demographic structure occurred at generation $t = 1500$. (c) The demographic structure occurred at generation $t = 4000$. (d) The demographic structure occurred at generation $t = 6000$. Adapted from the paper \cite{Liu17IIR}.}
  \label{fig:case3gra}
\end{figure*}

One example where cooperators and cheaters coexist is Fig. \ref{fig:case3gra}, the system with $n = 3$ and $u = 2$. When considering the demographic structure where $s_{12}$ and $s_{23}$ (with $s_{11}$) dominate with some fraction of $s_{22}$ (Fig. \ref{fig:case3gra_3}), we would say that $s_{22}$ is the cheater with respect to the pair of cooperators $s_{12}$ and $s_{23}$, because it does not produce $\bar{3}$ needed by the cooperator $s_{23}$, and only consumes $\bar{2}$, the metabolites needed by $s_{12}$ and $s_{23}$. From both simulations and equations in the paper \cite{Liu17IIR}, we know that $s_{22}$, $s_{12}$ and $s_{23}$ coexist. Even, the cheater $s_{22}$ prevents the system from collapsing. The possible benefits of cheaters to the whole system are also discussed in other works \cite{MacLean10AMO, Hummert14EGT}.

But wait a minute. Is it reasonable to consider $s_{22}$ as the cheater in the first place? Recall that it is the first resident in the system (Fig. \ref{fig:case3gra_1}), but because of the appearance of $s_{12}$ and then $s_{23}$, it is driven towards extinction. In fact, I would like to say that, in the beginning $s_{12}$ is the cheater since it produces $\bar{3}$ which no current resident needs. Moreover, the ``cooperator'' $s_{12}$ does not plan to cooperate in the beginning. Its appearance is just due to the fact that there are abundant resources. We should be more careful when we use the term: ``cheater''. 

The whole process can be interpreted as the following way. At first $s_{11}$ and $s_{22}$ cooperate to survive, constituting the demographic structure with two species coexisting (Fig. \ref{fig:case3gra_1}). Then $s_{12}$ grows as the cheater of the current demographic structure, and then $s_{23}$ grows as the cooperator of $s_{12}$. Finally $s_{12}$ and $s_{23}$ drive the original cooperator $s_{22}$ extinct and replace its ``niche'' (Fig. \ref{fig:case3gra_inter} - \ref{fig:case3gra_2}). On this stage, $s_{12}$ becomes the cooperator (not the cheater any more), cooperating with $s_{23}$ and $s_{11}$ to survive. Summarising it in general words, the cheater first invades a cooperative system, but afterwards the system adapts itself so that the cheater becomes the cooperator.

There is a real-world scenario possibly associated with this process. Corrinoid-dependent reactions are prevalent in various metabolic processes, but only a subset of prokaryotes produce corrinoids. In fact, around $75\%$ bacteria are supposed to encode corrinoid-dependent enzymes, but at least half of them cannot produce corrinoids \textit{de novo} \cite{Seth14NCI}. In Fig. \ref{fig:case3gra}, if we consider $\bar{1}$ as the corrinoid, the ratio of corrinoid producers over consumers becomes much smaller as the system evolves from Fig. \ref{fig:case3gra_1} to \ref{fig:case3gra_2}. For the former demographic structure, the producer is $s_{22}$ and the consumer is $s_{11}$, and the ratio of population is $2/1$; while for the latter, the producer is $s_{23}$, and consumer is $s_{11}$ and $s_{12}$, and the ratio is $2/3$.

So the story may be as following. In the beginning, corrinoids are produced by most of the microbes, and leaked into the system. Afterwards, corrinoids become so abundant that the microbes without the ability to produce them can easily get them from the environment. So the capable microbes have no advantage over these incapable ones, namely ``cheaters''. Since more gene means more burden, the redundant capable microbes would die so that it reaches an equilibrium (maybe a ``meta-equilibrium'') demographic structure where ``cooperators'' and ``cheaters'' coexist, and the level of corrinoids is kept at a reasonable level.

\section{Future works}
\label{sec:future}

In this section, I list intriguing points and questions related to the number soup model, in the order of relevance and importance, that deserve more reflections and investigations.

\begin{itemize}
\item Do all ecosystems collapse at the end?

In the number soup model, system collapses are not rare, and the existence of keystone species is quite common. So the question above is naturally raised, although we may think that some systems could be able to persist for a very long time. In the context of the model, this question can be answered by systematically investigating the statistics of time till collapse.

\item It seems that mutation and biodiversity help prolong the whole system.

In the number soup model, we observed that mutation and biodiversity help prolong the whole system. That is because the extinct species could be possibly regenerated by mutations. In Bell's generalised predator-prey model \cite{Bell2007TEO}, it was also observed that mutations prolong the existence of trophically complex communities. However, in the context of natural ecosystems, mutations are not quick enough to regenerate the extinct keystone species or similar species. Instead, the role played by mutations could be replaced by biodiversity. That is, if there is rich biodiversity, some other species may take over the vacant niche just appeared due to the extinction of the keystone species, or the ecologically extinct keystone species is reintroduced.

\item It seems that natural selection does not maximise stability.

In the number soup model, there is no evidence showing that the system tends to evolve to more and more stable demographic structure. Selection only works on the individual level so that the organisms which are more efficient in obtaining metabolites are selected. This does not automatically lead to a maximisation of the stability of the whole system. In other words, natural selection seems not necessarily maximise stability. It was also observed in Bell's predator-prey model \cite{Bell2007TEO}.

On the other hand, in Gedeon's work \cite{Gedeon2015DOS}, for a given set of species surviving, there is at most one equilibrium, and there is unique stable equilibrium which corresponds to the one having the greatest number of species, supported by the available resources. From the number soup model, I get a different opinion about the stability of ecosystems: There is no ultimately equilibrium for the whole system.

\item It seems that ecosystems evolve towards robust yet fragile.

In the number soup model, the ecosystem always evolves towards more and more ``meta-stable'' demographic structures, in the sense that very unstable demographic structures cannot persist in a relatively long period of time and they always vanish very quickly. On the other hand, the meta-stable demographic structures are always sensitive to certain kinds of perturbations. That is to say, the ecosystem inevitably evolves towards a robust yet fragile state.

\item The number soup model supports the Gaia hypothesis and the maximum power principle.

At a much larger scale, Earth can be considered as a closed system with respect to its chemical components, with the only inflow of the sunlight, and the biosphere self-organises to efficiently consume the sunlight at a large time scale (here ``consume'' means radiating the energy from the sun in the form of heat). This leads to the Gaia hypothesis \cite{Lovelock1974AHB}. This property of biosphere is very similar with what is observed in the ecosystem of the number soup model.

On the other hand, there was an unfamous theory, the maximum power principle, proposed by Howard Thomas Odum, saying that during self-organisation, system maximises power intake, and uses those power to reinforce production and efficiency. This is quite similar with what has been observed in the number soup model. In addition, in Gedeon's model of resource-consumer interactions \cite{Gedeon2015DOS}, the equilibrium of the ecosystem also maximises utilisation of available resources.

\item Resource recycling seems a common property of self-organised ecosystems.

Constantly consuming up the inflow means that the system is in balance dynamically. So substances in the system have to be recycled. In the number soup model, the metabolites left in the system are frequently consumed and produced by organisms, acting as ``catalysts'', i.e., they are recycled. In natural cases, for sludge granules (or guts microbiota), the metabolites and substances contained in granules (or the substances in the guts of the host animal) are recycled \cite{Liu17IIR}; while for Earth, various elements are recycled \cite{Crombach2009EOR}, driven by the process that the energy is frequently stored in chemical bonds in the form of chemical energy, and released in the form of heat. Besides natural ecosystems, Crombach and Hogeweg have also observed that the digital organisms in their model are adapted to recycle resources spontaneously \cite{Crombach2009EOR}. Resource recycling might be a common property of self-organised ecosystems.

\item State-switching of ecosystems is observed both in nature and in the number soup model.

Many ecosystems such as lakes, coral reefs and forests, have alternative states, between which it could switch abruptly. Loss of resilience of the ecosystem, due to the smooth changing of external conditions such as temperature and nutrient loading, paves the way for abrupt switching \cite{Scheffer2001CSI}. After the loss of resilience, small internal or external stochastic perturbations could trigger the switching. It is not relevant to talk about the changing of external condition in the number soup model since the external condition is always constant here. However, even in the number soup model where it has a constant external condition, abrupt state-switching is still observed.

\item System collapse seems possible to be forecast by early-warning signals.

In the number soup model we sometimes observe large fluctuations right before complete system collapse, although large fluctuations are not necessarily associated with complete collapse. In other situations where the system abruptly switches to another demographic structure, sometimes there are also large fluctuations beforehand, which could be considered as incomplete collapse followed by recovery. Therefore, by observing whether fluctuations get unexpectedly large, or extracting some early-warning signals, we might be able to tell whether a collapse or transition would follow. These early-warning signals are also proposed to forecast whether a natural ecosystem is at the critical point \cite{Carpenter2011EWO, Scheffer2009ESF}.

\item Is it easier to go extinct with too high birth rate?

In the number soup model, the reason why resources are consumed so efficiently is that the birth rate of organisms is so high that the population always gets saturated, and then they fully consume the resources. In the real world, microbes consume resources and reproduce as much as possible, which is similar with the number soup model. But for larger organisms such as birds, mammals, etc., they seem not to reproduce as much as possible, or at least the resources would not be consumed as fast as microbes do.

Why do not large organisms fully consume resources? Superficially, it is due to the low net birth rate. But why do these large organisms have low net birth rate? Here are my guesses: (1) since large organisms are large resource consumers, if they reproduce too fast, they would consume up resources quickly and then die out, so the evolution selects large organisms with low birth rate; or (2) the low birth rate is just due to the physiological constraints, e.g., large bodies are too large and complex to reproduce faster. The first guess can be investigated by the number soup model by setting different birth and death rate, to see whether organisms with too high birth rate have high probabilities to die out. Alternatively, we can ask, which community, under-populated or over-populated one, is easier to go extinct?

From another point of view, high birth rate is beneficial for individual organism, but could be deleterious for the whole species, namely deleterious in a system level. So a question is raised: Is there a tradeoff of birth rate between individual level and system level?

\item The number soup model can be considered as a metaphor of financial markets.

The similarities between ecology and financial markets have been studied, of what is called ``financial ecosystem'' \cite{Haldane2011SRI, Smerlak2015MSR}. In ecology, the interactions are trophic interactions among organisms; while in financial markets, they are typically debit and credit relationships among banks, or industrial chains. The number soup model can be considered as a metaphor of financial markets. The metabolites are the resources (and money) in circulation. The organisms are the industries and/or banks, due to the facts that (1) both of them reproduce (or expand in the financial case) if they take in metabolites (or the resources and money in the financial case), and (2) in both cases, individuals depend on each other.

We observed some common phenomena in both systems. System collapse is one of them. In the number soup model, the system collapse is caused by the lack of keystone species and some certain metabolites in circulation. In real financial markets, some people claim that catastrophic decline is caused by lack of liquidity \cite{Acemoglu2013SRA}, which is the counterpart of metabolites in circulation in the number soup model.

\item More immature questions.

What role does the parameter $a$ play in the number soup model? Do species loops with more species have advantages over the ones with less species? Do all ecosystems evolve to ultimately act as a black body?

\end{itemize}

%

\pnasbreak

\bibliography{ms.bbl}

\end{document}